\documentclass[aps,pre,groupedaddress,twocolumn,10pt]{revtex4-1}
\usepackage{graphicx}
\usepackage[english]{babel}
\usepackage[T1]{fontenc}

\usepackage{amsmath}
\usepackage{amssymb}
\usepackage{amsthm}
\usepackage{bbm}
\usepackage{hyperref}
\usepackage[usenames,dvipsnames]{xcolor}
\hypersetup{colorlinks=true, linkcolor=BrickRed, urlcolor=blue!50!black, citecolor=blue!50!black}

\usepackage[normalem]{ulem}

\newcommand\diff{\mathrm{d}}
\renewcommand{\vec}[1]{\mathbf{#1}}
\renewcommand{\imath}[0]{\mathsf{i}}

\usepackage[nodayofweek,level]{datetime}

\begin{document}

\title{Bimodal probability density characterizes the elastic behavior of a semiflexible polymer in 2D under compression}
\author{Christina Kurzthaler$^1$ and Thomas Franosch$^{1}$}
\affiliation{$^1$Institut f\"ur Theoretische Physik, Universit\"at Innsbruck, Technikerstra{\ss}e 21A,
A-6020 Innsbruck, Austria}
\email{thomas.franosch@uibk.ac.at}

\begin{abstract}
We explore the elastic behavior of a wormlike chain under compression in terms of exact solutions for the associated probability densities.
Strikingly, the probability density
for the end-to-end distance projected along the applied force
exhibits a bimodal shape in the vicinity of the critical Euler buckling force of an elastic rod, reminiscent
of the smeared discontinuous phase transition of a finite system.
These two modes reflect the almost stretched and the S-shaped configuration of
a clamped polymer induced by the compression.
Moreover, we find a bimodal shape of the probability density for the transverse fluctuations of
the free end of a cantilevered polymer as fingerprint of its semiflexibility.
In contrast to clamped polymers, free polymers display a circularly symmetric probability density
and their distributions are identical for compression and stretching forces.
\end{abstract}

\date{\formatdate{15}{3}{2018}}

\maketitle

%%%MAIN TEXT%%%%
\section{Introduction}
Semiflexible polymers are abundant in nature and play a pivotal role for the mechanical
properties of single cells~\cite{Sackmann:1994}.
Examples of these biopolymers include filamentous actin~\cite{Lieleg:2010}, microtubules~\cite{Brangwynne:2006},
and intermediate filaments~\cite{Nolting:2014}, which all constitute integral parts of cells
and form the ingenious scaffold inside the cell, the cytoskeleton~\cite{Bausch:2006}.
Together, these filaments account for the cell shape and its ability to adapt dynamically to its
environment, its  mechanical and structural stability, cell motility,
intracellular transport processes, and also cell division~\cite{Brangwynne:2006, Fletcher:2010,Lieleg:2010}.
As the interior of a cell is densely packed and crowded with macromolecules,
the environment of these filaments becomes constrained and alters their elastic behavior drastically~\cite{Nolting:2014}.
Cytoskeletal polymers often form bundles or networks, which are cross-linked by a large variety of
regulatory proteins, or accumulate in entangled solutions~\cite{Amuasi:2015,Carillo:2013, Huisman:2008,Kroy:1996,MacKintosh:1995,Plagge:2016,Razbin:2015,Storm:2005}.
Already in the force-free case, single polymers experience stretching and compression forces induced by their
surrounding network, and consequently the distance between two ends of a polymer, that are, for instance,
fixed by cross-links, differs significantly form their contour length.
In the presence of mechanical stresses these networks exhibit fascinating nonlinear
behavior on larger, macroscopic scales~\cite{Amuasi:2015,Carillo:2013,Chaudhuri:2007,Claessens:2006,Huisman:2008,Kroy:1996,MacKintosh:1995,Plagge:2016,Razbin:2015,Storm:2005},
such as the stiffening of materials with increasing strain~\cite{Carillo:2013},
or the reversible stress-softening behavior of filamentous actin networks~\cite{Chaudhuri:2007}.

There has been significant progresss in the recent past to elaborate the principles for the peculiar nonlinear elastic behvior~\cite{MacKintosh:2014},
yet the mechanisms still require further elucidation at different levels of coarse-graining.
In particular, the macroscopic behavior strongly depends on its single components and
the mechanical properties of single filaments serve as an essential input to fully understand the
elasticity of networks~\cite{Hugel:2001}. Already single semiflexible polymers respond sensitively to
external forces and, thereby, exhibit peculiar stretching and bending behavior~\cite{Baczynski:2007, MacKintosh:2014,Magnasco:1993, Razbin:2016}.
Experimental methods, including optical~\cite{Ashkin:1997,Mehta:1999} and magnetic tweezers~\cite{Gosse:2002},
transmission electron microscopy~\cite{Kuzumaki:2006}, and acoustic~\cite{Sitters:2015} and atomic force
spectroscopy~\cite{Hugel:2001,Janshoff:2000}, have been applied to measure \emph{in vitro} the nonlinear force-extension relation
of purified biopolymers, such as DNA ~\cite{Bouchiat:1999,Bustamante:2000,Marko:1995,Sitters:2015} and actin filaments~\cite{Liu:2002},
single molecules, e.g. titin~\cite{Kellermayer:1997} and collagen~\cite{Sun:2002},
and also synthetic carbon nanotubes~\cite{Kuzumaki:2006}.
Yet, in striking contrast to flexible polymers, the probability distribution for the end-to-end distance of semiflexible polymers
is expected to deviate significantly from a simple Gaussian behavior~\cite{Wilhelm:1996},
and, thus, the mean end-to-end distance is not necessarily a good indicator for the shape of the associated probability distribution.
Consequently, the full probability distribution is required to obtain a more general
characterization of the elasticity of semiflexible polymers.
Experimentally, fluorescence videomicroscopy has already been applied to access
the probability distribution for the end-to-end distance of force-free actin filaments~\cite{LeGoff:2002}
and semiflexible polymers confined to microchannels~\cite{Koster:2005,Koster:2009},
but permits in principle also to extract reliable information on the polymer configurations
induced by external loads.

A widely used model to describe the elastic properties of semiflexible polymers is the celebrated wormlike
chain model, also referred to as Kratky-Porod model~\cite{Kratky:1949}.
The force-free behavior of a wormlike chain with free ends has been elaborated analytically for
the end-to-end probability density in the weakly-bending approximation~\cite{Wilhelm:1996} and,
furthermore, evaluated numerically for polymers of arbitrary stiffness using an inverse
(Fourier) Laplace transform~\cite{Samuel:2002, Mehraeen:2008}.
In addition, the probability density for the transverse fluctuations of the free end of a cantilevered
polymer has been extracted from computer simulations~\cite{Lattanzi:2004} and qualitatively confirmed
by an approximate theory~\cite{Benetatos:2005} in 2D and computed formally exactly in 3D by inverting
an infinite matrix~\cite{Semeriyanov:2007}.
To explore the response of semiflexible polymers to external forces, we have recently provided
exact expressions for the two lowest-order moments, namely, the force-extension relation
and the susceptibility (i.e. the variance)~\cite{Kurzthaler:2017},
thereby, complementing theoretical and simulation studies on the stretching behavior~\cite{Prasad:2005}
and the response of single semiflexible polymers
upon compression in the weakly-bending approximation~\cite{Kroy:1996,Baczynski:2007}.
The smearing of the classical Euler buckling instability by thermal fluctuations leads
to a rapid decrease of the force-extension relation and a strongly peaked susceptibility, reminiscent
of a smeared discontinuous phase transition in a finite system. To corroborate this scenario and to
fully understand the behavior of a semiflexible polymer under compression, access to the full probability
distribution for the end-to-end distance is required.

Here, we first provide analytic expressions for the characteristic function, i.e. the Fourier
transform of the probability density, of
force-free wormlike chains with clamped ends, one free and one clamped end (i.e. cantilevered), and
free ends reflecting different experimental setups.
The exact solutions then permit to evaluate numerically via an inverse Fourier transform
the associated probability densities for the end-to-end distance projected along or
transverse to the clamped ends of the polymer.
Furthermore, we explore the behavior of semiflexible polymers with different bending rigidities under compression and tension
and provide insights into the polymer configurations via
the probability density for the end-to-end distance projected onto the
direction of the applied force. We validate our analytic results with the mean force-extension
relation and the associated variance, elaborated previously~\cite{Kurzthaler:2017}.
Moreover, we have performed simulations for selected parameter sets and exemplarily compared them
to our exact theory.

\section{Wormlike chain model}
\begin{figure}[htp]
 \centering
\includegraphics[width = 0.9\linewidth]{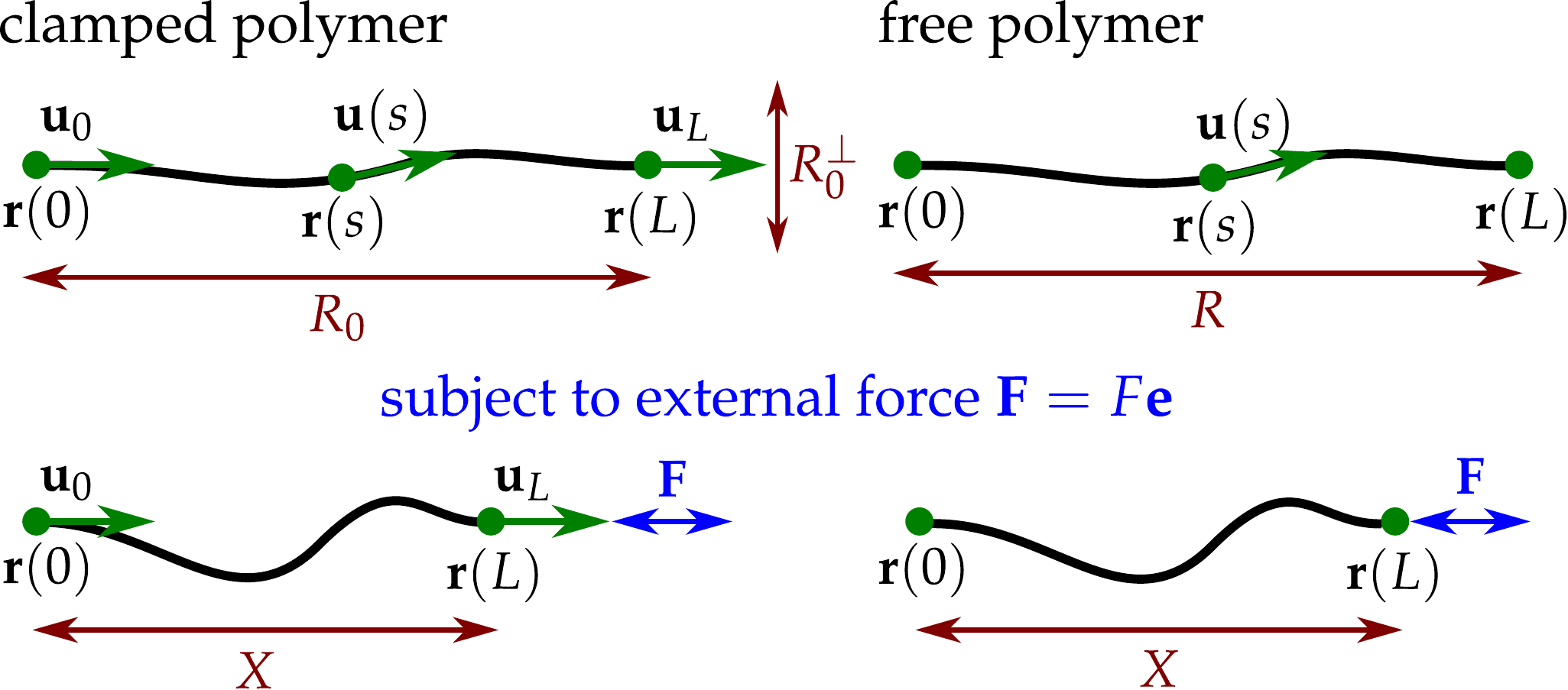}
\caption{Wormlike chain with clamped (left) and free ends (right). Here, $\vec{r}(s)$ is
the contour at arc length $s$, $\vec{u}(s)=\diff\vec{r}(s)/\diff s$ is the tangent vector
of the polymer along the contour of length $L$, $\vec{u}_0=\vec{u}_L$ are the clamped ends,
and $\vec{F}=F\vec{e}$ is the applied compression, $F<0$, or pulling, $F>0$, force along the fixed direction $\vec{e}$.
Moreover, $R_0=\int_0^L\diff s \ \vec{u}_0\cdot\vec{u}(s)$
denotes the end-to-end distance of a force-free polymer projected onto the clamped ends,
$R_0^\perp=\int_0^L\diff s \ \vec{u}_0^\perp\cdot\vec{u}(s)$ the transverse end-to-end distance,
$R=|\int_0^L \diff s \ \vec{u}(s)|$ the end-to-end distance of a free polymer, and
$X =\int_0^L\diff s \ \vec{e}\cdot\vec{u}(s)$ the end-to-end distance projected onto the direction of the applied force.
\label{fig:wlc}}
\end{figure}

To explore the elastic properties of a single semiflexible polymer
we employ the well-established \textit{wormlike chain model}  (WLC model).
Here, the bending energy of a wormlike chain~\cite{Kratky:1949} is
expressed by its squared curvature,
\begin{align}
  \mathcal{H}_0	&= \frac{\kappa}{2}\int_0^L\!\diff s \ \left(\frac{\diff \vec{u}(s)}{\diff s}\right)^2, \label{eq:wlc_h0}
\end{align}
where $s$ denotes the arc length of the polymer, $L$ its contour length, and $\kappa$ the bending rigidity.
Furthermore, $\vec{u}(s)=\diff\vec{r}(s)/\diff s$ is the tangent vector of the polymer along its contour $\vec{r}(s)$
with unit length, $|\vec{u}(s)|=1$, see Fig.~\ref{fig:wlc}.
In the framework of statistical physics the corresponding partition sum $Z_0(\vec{u}_L,L|\vec{u}_0,0)$
for a clamped polymer with initial orientation $\vec{u}_0$
and final orientation $\vec{u}_L$ is formally obtained as a
path integral of Boltzmann weights over all possible chain configurations,
which obey the local inextensibility constraint $|\vec{u}(s)|=1$,
\begin{align}
  Z_0(\vec{u}_L,L|\vec{u}_0,0)  &= \int _{\vec{u}(0)=\vec{u}_0}^{\vec{u}(L)=\vec{u}_L}\mathcal{D}[\vec{u}(s)]\exp\left(-\frac{\mathcal{H}_0}{k_\text{B}T}\right).\label{eq:Z0}
\end{align}
Here, $k_\text{B}$ denotes the Boltzmann constant and $T$ is the temperature of the system.
Although the Hamiltonian is quadratic the path integral cannot be solved by Gaussian integrals
due to the local inextensibility constraint. Yet, an exact solution of the partition sum
can be elabaroted by deriving the corresponding Fokker-Planck equation and solving it
in terms of associated eigenfunctions.
In particular, in 2D the partition sum can formally be expanded into Fourier modes~\cite{Spakowitz:2004},
which naturally introduces the persistence length $\ell_\text{p}=2\kappa/k_\text{B}T$ as the decay length of
the tangent-tangent correlations of the polymer. In 3D, where the solution of the partition sum
can be formally achieved in spherical harmonics, the persistence length is given by $\ell_\text{p}= \kappa/k_\text{B}T$.
The persistence length represents a geometric measure for
the stiffness of the polymer and allows discriminating between flexible, $\ell_\text{p}/L \ll 1$, stiff,
$\ell_\text{p}/L \gg 1$, and semiflexible polymers, $\ell_\text{p}/L \simeq 1$~\cite{MacKintosh:2014,Doi:1986}.

To elucidate the response of a semiflexible polymer to an external force $\vec{F}$,
we account for the stretching energy,
\begin{align}
 \mathcal{H}_\text{force} &= -\int_0^L\! \diff s \ \vec{F}\cdot\vec{u}(s),
\end{align}
where the force $\vec{F}=F\vec{e}$ acts along a fixed direction $\vec{e}$, $|\vec{e}|=1$,  and can be either tension, $F>0$, or compression, $F<0$.
Comparing the force to the classical Euler buckling force $F_c$ permits to distinguish regimes of small and strong compression forces.
The total energy of the system then reads
\begin{align}
 \frac{\mathcal{H}}{k_\text{B}T} &= \int_0^L \!\diff s \left[\frac{\kappa}{2k_\text{B}T}\left(\frac{\diff \vec{u}(s)}{\diff s}\right)^2-f\vec{e}\cdot\vec{u}(s)\right], \label{eq:hamiltonian}
\end{align}
where we have introduced the reduced force $f=F/k_\text{B}T$.
Since $f$ has units of an inverse length, a different choice of a dimensionless force parameter would be $fL$ or $f\ell_\text{p}$.
The corresponding partition sum,
\begin{align}
Z(\vec{u}_L,L|\vec{u}_0,0)  &= \int _{\vec{u}(0)=\vec{u}_0}^{\vec{u}(L)=\vec{u}_L}\mathcal{D}[\vec{u}(s)]\exp\left(-\frac{\mathcal{H}}{k_\text{B}T}\right),\label{eq:polymer_Z}
\end{align}
can be computed by solving the associated Fokker-Planck equation~\cite{Prasad:2005,Kurzthaler:2017},
\begin{align}
 \partial_s Z(\vec{u},s|\vec{u}_0,0) = \left[f\vec{e}\cdot\vec{u}+\frac{k_\text{B}T}{2\kappa}\Delta_\vec{u}\right]Z(\vec{u},s|\vec{u}_0,0),\label{eq:part_sum_D}
 \end{align}
 with $\Delta_\vec{u}$ the angular part of the Laplacian. It is subject to the initial condition $Z(\vec{u},s=0|\vec{u}_0,0)=\delta(\vec{u},\vec{u}_0)$, where
the $\delta$-function enforces both directions to coincide.
In particular, the Fokker-Planck equation describes the evolution of the partition sum of a polymer as its arc length $s$ increases,
given that the initial orientation of the polymer at $s=0$ is set by $\vec{u}(0)=\vec{u}_0$.

In 2D the orientation of the polymer, $\vec{u}=(\cos(\varphi),\sin(\varphi))^T$, can be parametrized in terms of the polar angle $\varphi=\angle(\vec{e},\vec{u})$,
which is measured here with respect to the applied force. Thus, the Fokker-Planck equation for the partition sum $Z(\varphi,s|\varphi_0,0)$ reads
\begin{align}
\partial_s Z(\varphi,s|\varphi_0,0) &= \left[f\cos(\varphi)+\frac{1}{\ell_\text{p}}\partial_\varphi^2\right] Z(\varphi,s|\varphi_0,0).\label{eq:part_sum_f}
\end{align}
This equation is reminiscent of the Schr\"odinger equation of a quantum pendulum~\cite{Aldrovandi:1980} and
evaluates to an expansion in even and odd Mathieu functions, as has been
elaborated for stretching forces, $f>0$, in Ref.~\cite{Prasad:2005} and for compression forces, $f<0$, in Ref.~\cite{Kurzthaler:2017}, respectively.

In the following sections we provide a theoretical framework to obtain the probability density for
the end-to-end distance  projected along or transverse to the
clamped ends of force-free semiflexible polymers.
Furthermore, we obtain the probability density for the projected end-to-end distance of
a semiflexible polymer subject to external forces. We discuss three different experimental setups
encoded in the boundary conditions.

\subsection{Probability density for the (projected) end-to-end distance of a force-free semiflexible polymer}
We first discuss in detail the computation of the characteristic functions of clamped, cantilevered, and
free semiflexible polymers, which fully characterize the probability densities of the projected end-to-end distance.
We derive the corresponding probability densities by an inverse Fourier transform of the characteristic
functions.
\subsubsection{Clamped polymer}
Here, we derive the probability density $\mathbb{P}(R_0|\vec{u}_L,\vec{u}_0)$
for the end-to-end distance
projected along a fixed direction $\vec{e}$ of a clamped polymer, $R_0 = \int_0^L \diff s \ \vec{e}\cdot\vec{u}(s)$,
where the initial and final orientations, $\vec{u}_0$ and $\vec{u}_L$, are considered as fixed boundary conditions. It is defined by
\begin{align}
\begin{split}
\mathbb{P}(R_0|&\vec{u}_L,\vec{u}_0)= \left\langle\delta\!\left(R_0-\int_0^L \! \diff s \ \vec{e}\cdot\vec{u}(s)\right)\right\rangle_{0,\text{ BC}}\\
&= \frac{1}{Z_0}\int_{\vec{u}(0)=\vec{u}_0}^{\vec{u}(L)=\vec{u}_L}\mathcal{D}[\vec{u}(s)]\exp\left(-\frac{\mathcal{H}_0}{k_\text{B}T}\right)\\
& \ \ \ \ \qquad \qquad \times\delta\!\left(R_0-\int_0^L \! \diff s \ \vec{e}\cdot\vec{u}(s)\right),
\end{split}
\end{align}
where $\langle\cdot\rangle_{0,\text{ BC}}$ denotes the average with respect to the Boltzmann distribution with Hamiltonian
$\mathcal{H}_0$ [Eq.~\eqref{eq:wlc_h0}] and normalization $Z_0\equiv Z_0(\vec{u}_L,L|\vec{u}_0,0)$ [Eq.~\eqref{eq:Z0}] such that
the ends fulfill the boundary conditions (BC) $\vec{u}(0)=\vec{u}_0$ and $\vec{u}(L)=\vec{u}_L$.
To emphasize that the orientations $\vec{u}_0$ and $\vec{u}_L$ now serve as boundary conditions, rather than initial and final values
as in Eqs.\eqref{eq:part_sum_D} and \eqref{eq:part_sum_f}, we write the probability as conditioned with respect to the boundary conditions.
The associated characteristic function for the wavevector $\vec{k}=k\vec{e}$ along the fixed direction $\vec{e}$
is defined as the Fourier transform,
\begin{align}
 \widetilde{\mathbb{P}}(k|\vec{u}_L,\vec{u}_0) 	&= \int_{-\infty}^\infty\!\diff R_0 \ \exp(-\imath k R_0)\mathbb{P}(R_0|\vec{u}_L,\vec{u}_0).\label{eq:fourier_transform_prob}
\end{align}
It can be rewritten
\begin{align}
\begin{split}
\widetilde{\mathbb{P}}(k|\vec{u}_L,\vec{u}_0) &=\frac{Z_0(k,\vec{u}_L,L|\vec{u}_0,0)}{Z_0(\vec{u}_L,L|\vec{u}_0,0)}\\
						&= \left\langle \exp\left[-\imath k\int_0^L\!\diff s \ \vec{e}\cdot\vec{u}(s)\right] \right\rangle_{0,\text{ BC}},
\end{split}
\end{align}
where we have abbreviated the path integral by the partition sum for the wavenumber~$k$
\begin{align}\begin{split}
 &Z_0(k,\vec{u}_L,L|\vec{u}_0,0) =\\
&= \int^{\vec{u}(L)}_{\vec{u}(0)}\!\mathcal{D}\left[\vec{u}(s)\right]\exp\left[-\left(\frac{\mathcal{H}_0}{k_\text{B}T}+\imath k \int_0^L\!\diff s \ \vec{e}\cdot\vec{u}(s)\right)\right].\label{eq:characteristic_polymer_int}
\end{split}
\end{align}
The path integral differs from the partition sum of a polymer subject to an external force, $Z(\vec{u}_L,L|\vec{u}_0,0)$ [Eq.~\eqref{eq:polymer_Z}],
only by allowing the force to become formally complex.
Therefore, the Fokker-Planck equation associated with the path integral assumes the same form,
via the mapping $f\mapsto -\imath k$,
\begin{align}
 \partial_s &Z_0(k,\varphi,s|\varphi_0,0)= \notag\\
 &=\left(-\imath k \cos(\varphi) +\frac{1}{\ell_\text{p}}\partial^2_\varphi\right)Z_0(k,\varphi,s|\varphi_0,0).\label{eq:characteristic_polymer}
\end{align}
Here, we have employed again the polar angle $\varphi=\angle(\vec{e},\vec{u})$ between the direction $\vec{e}$
and the tangent vector~$\vec{u}$.
Thus, the equation can again be solved using an expansion in appropriate angular eigenfunctions $z(\varphi)$. Inserting the separation ansatz $\exp(-\lambda s)z(\varphi)$
into Eq.~\eqref{eq:characteristic_polymer} we obtain the eigenvalue problem
\begin{align}
 \left[\lambda -\imath k\cos(\varphi)+\frac{1}{\ell_\text{p}}\frac{\diff^2}{\diff \varphi^2}\right]z(\varphi) &=0.
\end{align}
A change of variables, $x=\varphi/2$, and rearranging terms leads to the well-known Mathieu equation~\cite{NIST:online,NIST:print}
\begin{align}
 \left[\frac{\diff^2}{\diff x^2} +(a-2q\cos(2x))\right]z(x)&=0, \label{eq:mathieu}
\end{align}
with (yet imaginary) deformation parameter $q=2\imath k\ell_\text{p}$ and eigenvalue $a=4\ell_\text{p}\lambda$.

The general solution of Eq.~\eqref{eq:characteristic_polymer} is then expressed as a
linear combination of $\pi$-periodic even and odd Mathieu functions, $\text{ce}_{2n}(q,x)$ and $\text{se}_{2n+2}(q,x)$,
with associated eigenvalues $a_{2n}(q)=4\ell_\text{p}\lambda_n$ and
$b_{2n+2}(q)=4\ell_\text{p}\lambda_n$~\cite{NIST:online,NIST:print}, respectively.
The Mathieu functions constitute a complete, orthogonal, and normalized set
of eigenfunctions, $\int_{0}^{2\pi}\!\diff x \ \text{ce}_{2n}(q,x)\text{ce}_{2m}(q,x)=\delta_{nm}\pi$,
and similarly for $\text{se}_{2n+2}(q,x)$~\cite{Ziener:2012}.
As the Mathieu functions are periodic, they can be expressed as deformed cosines and sines:
\begin{align}
\text{ce}_{2n}(q,x)&=\sum_{m=0}^\infty A_{2m}^{2n}(q)\cos(2mx), \label{eq:mathieu_even}\\
\text{se}_{2n+2}(q,x)&=\sum_{m=0}^\infty B_{2m+2}^{2n+2}(q)\sin[(2m+2)x], \label{eq:mathieu_odd}
\end{align}
where the Fourier coefficients, $A_{2m}^{2n}(q)$ and $B_{2m+2}^{2n+2}(q)$, are fully determined
by the recurrence relations [Eqs.~\eqref{eq:rec_A}-~\eqref{eq:rec_B} in the appendix~\ref{sec:appendix_numerics}].

Hence, the full solution for the partition sum  in terms of the eigenfunctions reads
\begin{align}
\begin{split}
&Z_0(k,\varphi_L,L|\varphi_0,0)=\\
&= \frac{1}{2\pi}\sum_{n = 0}^\infty\bigl[\text{ce}_{2n}(q,\varphi_0/2)\text{ce}_{2n}(q,\varphi_L/2)e^{-a_{2n}(q)L/4\ell_\text{p}}\\
    &  \ \ \ \  +\text{se}_{2n+2}(q,\varphi_0/2)\text{se}_{2n+2}(q,\varphi_L/2)e^{-b_{2n+2}(q)L/4\ell_\text{p}}\bigr]. \label{eq:characteristic_polymer_mathieu}
\end{split}
\end{align}
In particular, if $\varphi_0=0$ ($\varphi_0=\pi/2$ ) the wavevector is parallel (perpendicular)
to the initial orientation of the polymer.
Thus, our solution permits to recover the probability density for the end-to-end distance of the polymer
projected along the clamped ends or in any other direction.
The cases for clamped polymers follow by setting $\varphi_L=\varphi_0$.

Interestingly, the conformations of the semiflexible polymer can be regarded as the trajectory of a
self-propelled particle, which suggests that the methodology of both systems is intimately related
to one another. In particular, exact solutions for the partition sum of a semiflexible polymer [Eq.~\eqref{eq:characteristic_polymer_mathieu}]
can be directly mapped to the characteristic function of the displacements
of an active Brownian particle~\cite{Kurzthaler:2016,Kurzthaler:2017:circle,Kurzthaler:2017:janus}.
Thus, as for the mathematical analog of the self-propelled agent,
the eigenvalue problem [Eq.~\eqref{eq:mathieu}] is non-hermitian and, therefore,
the Mathieu functions and the corresponding eigenvalues assume complex values~\cite{Ziener:2012}.
The characteristic function of a clamped polymer becomes complex, and although it is formally the same as the partition sum of a
polymer subject to an external force [Eq.~\eqref{eq:polymer_Z}]~\cite{Kurzthaler:2017}, it exhibits a qualitatively different behavior.
For details on the numerical evaluation of the eigenfunctions and eigenvalues we refer to appendix~\ref{sec:appendix_numerics}.

The probability density for the projected end-to-end distance of a clamped polymer is obtained by an inverse Fourier transform
of the characteristic function,
\begin{align}
\mathbb{P}(R_0|\vec{u}_L,\vec{u}_0) &= \int_{-\infty}^\infty\!\frac{\diff k}{2\pi} \ \exp(\imath k R_0)\widetilde{\mathbb{P}}(k|\vec{u}_L, \vec{u}_0), \label{eq:polymer_dist_sol_c}
\end{align}
which we evaluate numerically using the Filon trapezoidal scheme~\cite{Tuck:1967}.

\subsubsection{Cantilevered polymer}
The probability density of a cantilevered polymer, $\mathbb{P}(R_0|\vec{u}_0)$,
can be obtained by marginalizing the probability density of a clamped polymer, $\mathbb{P}(R_0|\vec{u}_L,\vec{u}_0)$, i.e. integrating
over the final orientations, $\mathbb{P}(R_0|\vec{u}_0) = \int\!\diff \vec{u}_L \ \mathbb{P}(R_0|\vec{u}_L,\vec{u}_0)\mathbb{P}(\vec{u}_L|\vec{u}_0)$.
The probability density that a polymer has a final orientation $\vec{u}_L$ given the initial orientation $\vec{u}_0$,
reads $\mathbb{P}(\vec{u}_L|\vec{u}_0) = Z_0(\vec{u}_L,L|\vec{u}_0,0)/Z_0(L|\vec{u}_0,0)$, where the normalization evaluates to one,
$Z_0(L|\vec{u}_0,0)=\int\!\diff\vec{u}_L \ Z_0(\vec{u}_L,L|\vec{u}_0,0)=1$.
Then the corresponding characteristic function can be calculated by
\begin{align}
\begin{split}
\widetilde{\mathbb{P}}(k|\vec{u}_0)  	&= Z_0(L|\varphi_0,0)^{-1} \int_0^{2\pi}\!\diff \varphi_L \ Z_0(k,\varphi_L,L|\varphi_0,0) \\
					&= \sum_{n=0}^\infty A_0^{2n}(q)\text{ce}_{2n}(q,\varphi_0/2)\exp\left[-a_{2n}(q)L/4\ell_\text{p}\right],\label{eq:characteristic_polymer_hc}
\end{split}
\end{align}
where we have integrated Eq.~\eqref{eq:characteristic_polymer_mathieu} over the final orientations $\varphi_L$.
Here, $A_0^{2n}(q)$ are the Fourier coefficients of the even Mathieu functions [Eq.~\eqref{eq:mathieu_even}], while the odd Mathieu functions
do not contribute anymore.
As for a clamped polymer, the probability density $\mathbb{P}(R_0|\vec{u}_0)$ can be obtained numerically by an inverse Fourier transform~\cite{Tuck:1967} of
$\widetilde{\mathbb{P}}(k| \vec{u}_0)$.
\begin{figure*}[t]
 \centering
\includegraphics[width = \linewidth]{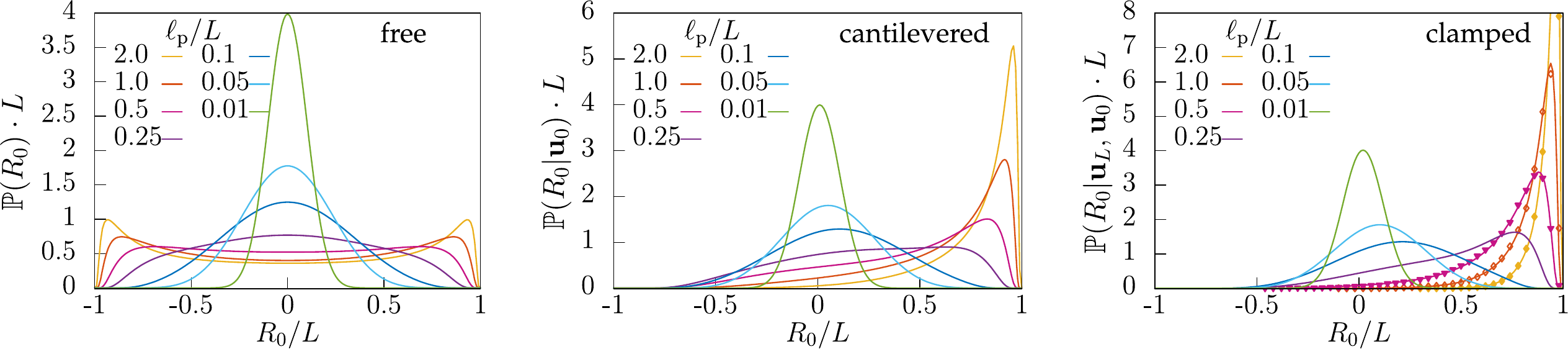}
\caption{Probability density for the end-to-end distance $R_0$ of a semiflexible polymer with free, $\mathbb{P}(R_0)$ (\textit{left panel}),
cantilevered, $\mathbb{P}(R_0|\vec{u}_0)$ (\textit{middle panel}),
and clamped ends, $\mathbb{P}(R_0|\vec{u}_L,\vec{u}_0)$ (\textit{right panel}), projected onto the
direction $\vec{e}$. For cantilevered and clamped polymers $\vec{e}$ is parallel to the initial (and final) orientation $\vec{u}_0$($=\vec{u}_L$), respectively.
Here, $L$ denotes the contour length and $\ell_\text{p}$ the persistence length of the polymer.
Selected pseudo-dynamic simulations are indicated by symbols.
\label{fig:prob_density_eq}}
\end{figure*}

\subsubsection{Free polymer}
The probability density, $\mathbb{P}(R_0)$, for the projected end-to-end distance $R_0$ of a free polymer is obtained by integrating the probability
density of a clamped polymer over the final and initial orientations,
$\mathbb{P}(R_0)=\int\!\diff\vec{u}_0 \int\!\diff\vec{u}_L \ \mathbb{P}(R_0|\vec{u}_L,\vec{u}_0)\mathbb{P}(\vec{u}_L|\vec{u}_0)\mathbb{P}(\vec{u}_0)$,
where the probability for the initial orientation is $\mathbb{P}(\vec{u}_0) = Z_0(L|\vec{u}_0,0)/Z_0(L)$ with normalization
$Z_0(L)=\int\!\diff\vec{u}_0 \ Z_0(L|\vec{u}_0,0)=2\pi$.
Then the Fourier transform evaluates to
\begin{align}
\begin{split}
 \widetilde{\mathbb{P}}(k) &=\frac{1}{Z_0(L)}\int_0^{2\pi}\!\diff \varphi_0\int_0^{2\pi}\!\diff \varphi_L \ Z_0(k,\varphi_L,L|\varphi_0,0)\\%= \left\langle \exp\left[-\imath \vec{k}\cdot\int_0^L\diff s \vec{u}(s)\right] \right\rangle_{0},
			      &= \sum_{n=0}^\infty\left[A_0^{2n}(q)\right]^2\exp\left[-a_{2n}(q)L/4\ell_\text{p}\right]. \label{eq:characteristic_polymer_free}
\end{split}
\end{align}
As for clamped and cantilevered polymers, the probability density $\mathbb{P}(R_0)$ for the projected end-to-end distance $R_0$
can be calculated numerically by an inverse Fourier transform~\cite{Tuck:1967} of
$\widetilde{\mathbb{P}}(k)$.

In addition to the probability density for the projected end-to-end distance, our method
permits to obtain an exact solution for the probability density, $\mathbb{P}(R)$, of the end-to-end distance,
$R=|\vec{R}|=|\int_0^L\!\diff s \ \vec{u}(s)|$, of a free polymer.
The probability density for the end-to-end distance $R$
of a free polymer is circularly symmetric and therefore
the associated characteristic function depends only on the magnitude of
the wavevector $k=|\vec{k}|$.
Thus, after averaging over the directions of $\vec{k}$,
we obtain the probability density by an inverse Hankel transform of the characteristic function $\widetilde{\mathbb{P}}(k)$
[Eq.~\eqref{eq:characteristic_polymer_free}] via
\begin{align}
  \mathbb{P}(R) 	&=  \int_0^\infty\! \frac{\diff k}{2\pi} \ kJ_0(kR)\widetilde{\mathbb{P}}(k), \label{eq:prob_density_free}
\end{align}
where $J_0(\cdot)$ denotes the Bessel function  of order zero. For the numerical evaluation we employ a Filon trapezoidal scheme~\cite{Barakat:2000}.

\subsection{Probability density for the end-to-end distance projected  onto the
applied force}
To include an external force into the theory developed in the previous section, in principle, we only have to
replace the Hamiltonian $\mathcal{H}_0$ [Eq.~\eqref{eq:wlc_h0}] by
$\mathcal{H}=\mathcal{H}_0+\mathcal{H}_\text{force}$ [Eq.~\eqref{eq:hamiltonian}], and adjust all quantities,
such as the normalization $Z_0(\vec{u}_L,L|\vec{u}_0)$ and the path integral $Z_0(k,\vec{u}_L,L|\vec{u}_0)$, accordingly.
Here, we restrict our discussion to forces $\vec{F}=F\vec{e}$ that are (anti-) parallel to the wave vector ($\vec{k}=k\vec{e}$),
depending on whether a compressive, $F<0$, or pulling force, $F>0$, is applied.

The solution strategy for the probability density,
\begin{align}
 \mathbb{P}(X|\text{BC})  &= \left\langle\delta\!\left(X-\int_0^L\! \diff s \ \vec{e}\cdot\vec{u}(s)\right)\right\rangle_{\text{BC}}, \label{eq:polymer_prob_force_u0}
\end{align}
remains the same as for the force-free distribution.
Note, that $X$ is always measured as end-to-end distance projected onto the
fixed direction $\vec{e}$. In fact, for clamped and cantilevered polymers it corresponds to the end-to-end distance
projected onto the direction of the applied force, whereas for free polymers
the end-to-end distance $X$ is measured in one specific direction $\vec{e}$ of the circularly symmetric probability density.

In principle, one can also derive the probability density by reweighting the force-free probability density
for the (projected) end-to-end distance $R$ ($R_0$)
with the Boltzman factor for the force bias, e.g. for clamped ends: $\mathbb{P}(X|\vec{u}_L,\vec{u}_0)\propto \exp(FX/k_\text{B}T)\mathbb{P}(R_0=X|\vec{u}_L,\vec{u}_0)$.
Yet, the numerical evaluation becomes unstable for compression forces in the vicinity of the critical Euler buckling force,
where the Boltzman factor exponentially inflates small round-off and truncation errors in the numerical evaluation of the
infinite sums [Eqs.~\eqref{eq:characteristic_polymer_mathieu},
\eqref{eq:characteristic_polymer_hc}, \eqref{eq:characteristic_polymer_free}].
Therefore, we rely on exact solutions for the corresponding characteristic functions.

The partition sums for clamped, $Z(\varphi_L,L|\varphi_0,0)$, cantilevered,
$Z(L|\varphi_0,0)$, and free polymers, $Z(L)$,
subject to external forces have been computed in
Refs.~\cite{Prasad:2005, Kurzthaler:2017}.
These are now used to normalize the characteristic functions (instead of $Z_0$).
To compute the path integral for the wave vector, $\vec{k}=k\vec{e}$, of a polymer subject
to an external force,
\begin{align}\begin{split}
 &Z(k,\varphi_L,L|\varphi_0,0) =\\
&=\int^{\vec{u}(L)}_{\vec{u}(0)}\!\mathcal{D}\left[\vec{u}(s)\right]\exp\left[-\left(\frac{\mathcal{H}_0}{k_\text{B}T}+\left(\imath k-f\right)\int_0^L\!\diff s \ \vec{e}\cdot\vec{u}(s)\right)\right],\label{eq:polymer_Z0_k}
\end{split}
\end{align}
we solve the associated Fokker-Planck equation,
\begin{align}
 \partial_s&  Z(k,\varphi,L|\varphi_0,0)= \notag\\
 &=\left[\left(f-\imath k\right) \cos(\varphi) +\frac{1}{\ell_\text{p}}\partial^2_\varphi\right]  Z(k,\varphi,L|\varphi_0,0),\label{eq:characteristic_polymer_force}
\end{align}
where $\varphi=\angle(\vec{e},\vec{u})$ is the angle between the direction $\vec{e}$ of the force/wavevector and the orientation $\vec{u}$,
and  $f=F/k_\text{B}T$ denotes the reduced force with units of an inverse length.
For compression forces, $f=-|f|$, the Fokker-Planck equation assumes the form of Eq.~\eqref{eq:characteristic_polymer}
with corresponding solution [Eq.~\eqref{eq:characteristic_polymer_mathieu}], yet,
with a different complex deformation parameter
\begin{align}
 q = 2\ell_\text{p}\left(|f|+\imath k\right).
\end{align}
Similarly, the characteristic functions for cantilevered and free polymers
can be transferred from Eq.~\eqref{eq:characteristic_polymer_hc} and from Eq.~\eqref{eq:characteristic_polymer_free}, respectively.
Interestingly, for pulling forces, $f = |f|$, a change of variables, $\varphi_0\mapsto \pi-\varphi_0$ and $\varphi_L\mapsto \pi-\varphi_L$,
in Eqs.~\eqref{eq:characteristic_polymer_hc} and~\eqref{eq:characteristic_polymer_mathieu}, and $q=2\ell_\text{p}\left(|f|-\imath k\right)$
yields the corresponding characteristic functions for cantilevered and clamped polymers, respectively.
For free polymers, the solution remains identical to that of a
compressive force [Eq.~\eqref{eq:characteristic_polymer_free}].

The probability densities for the end-to-end distance $X$ [Eq.~\eqref{eq:polymer_prob_force_u0}]
of a clamped, $\mathbb{P}(X|\varphi_L,\varphi_0)$,
cantilevered, $\mathbb{P}(X|\varphi_0)$, and free polymer, $\mathbb{P}(X)$,
are obtained by a numerical inverse Fourier transform of the corresponding characteristic functions~\cite{Tuck:1967}.

\section{Results -- force-free behavior}
To explore the response of a single semiflexible polymer to external forces, a full characterization
of the force-free polymer behavior serves as a reference.
We elucidate the force-free behavior of a wormlike chain with clamped and cantilevered
in terms of the probability densities for the projected end-to-end distance, as the clamping
introduces a characteristic direction to the system. Here, we focus on the projection
along and perpendicular to the clamped ends. We compare these results with the probability
density for the projected end-to-end distance of a free polymer and we
also discuss the probability density for the end-to-end distance of a free polymer,
as has been elaborated earlier~\cite{Samuel:2002, Mehraeen:2008}.
Exemplarily, we have corroborated selected exact predictions for the probability densities
with pseudo-dynamic simulations elaborated in Ref.~\cite{Kurzthaler:2017}
(see appendix~\ref{sec:appendix_simulations} for simulation details).

\subsection{Projected end-to-end distance (along clamped ends)}
Clamped ends naturally introduce a characteristic direction to the configurations of the polymers and, in contrast to
the circularly symmetric probability density of
a free polymer (see Fig.~\ref{fig:prob_density_eq} (\textit{left panel})), their probability densities projected along the clamped ends are asymmetric
(see Fig.~\ref{fig:prob_density_eq} (\textit{middle panel})  for cantilevered and Fig.~\ref{fig:prob_density_eq} (\textit{right panel})
for clamped polymers).
In particular, the asymmetry indicates that the configurations aligned along the opposite direction than the initial (and final)
orientations of the polymer remain less likely than the alignment along the clamped ends. This effect becomes most pronounced
for stiff polymers, where the most favorable end-to-end distance is almost their contour length, $R_0/L=1$.

However, the probability densities for the projected end-to-end distance reveal very similar
behavior for clamped and cantilevered polymers.
They exhibit a left-skewed form, i.e. the left
tail of the distribution is longer. The peak decreases
and the probability density broadens for decreasing stiffness due to thermal fluctuations.
For more flexible polymers an alignment into the opposite direction of the clamped ends becomes more probable
as the boundary conditions are less restrictive,
and, in particular, for flexible polymers, $\ell_\text{p}/L\ll 1$, we find a Gaussian distribution located at zero end-to-end distance,
which narrows for increasing flexibility.

\subsection{End-to-end distance of a free polymer\label{sec:force-free_free}}

\begin{figure}[h]
\centering
\includegraphics[width = 0.8\linewidth]{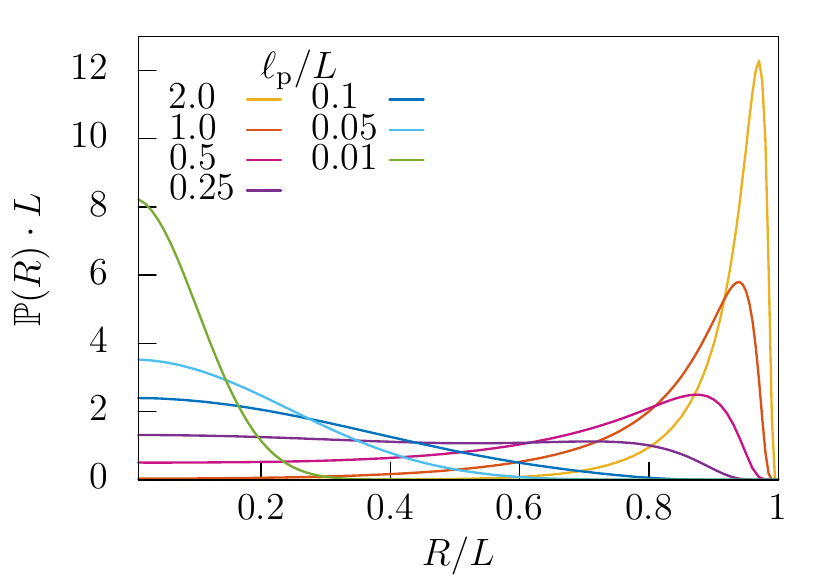}
\caption{Probability density $\mathbb{P}(R)$ for the end-to-end distance $R$ of a semiflexible polymer
with free ends. Here, $L$ denotes the contour length and $\ell_\text{p}$ the persistence length of the polymer.
\label{fig:prob_density_free}}
\end{figure}

The circularly symmetric end-to-end probability density of a free polymer displays peculiar behavior with
respect to the persistence length, see Fig.~\ref{fig:prob_density_free}.
For stiff polymers, the distribution exhibits a peak
located in the vicinity of a fully stretched configuration, $R/L=1$. Increasing the flexibility, shifts the
peak away, and a second peak, positioned close to zero, evolves at the transition between rather stiff and flexible polymers.
In particular, in the regime of $\ell_\text{p}/L\in[0.24,0.35]$
the distribution displays a bimodal shape, which is hardly visible
in the figure due to scaling.
For flexible polymers, $\ell_\text{p}/L\ll 1$, the peak approaches zero end-to-end distance,
as anticipated by the Gaussian chain. Here, the distribution becomes more narrow for more flexible polymers.
As expected, these findings are in agreement with earlier results obtained in Refs.~\cite{Samuel:2002, Mehraeen:2008}.

\subsection{End-to-end distance transverse to the clamped ends}

\begin{figure}[h]
\centering
\includegraphics[width = 0.8\linewidth]{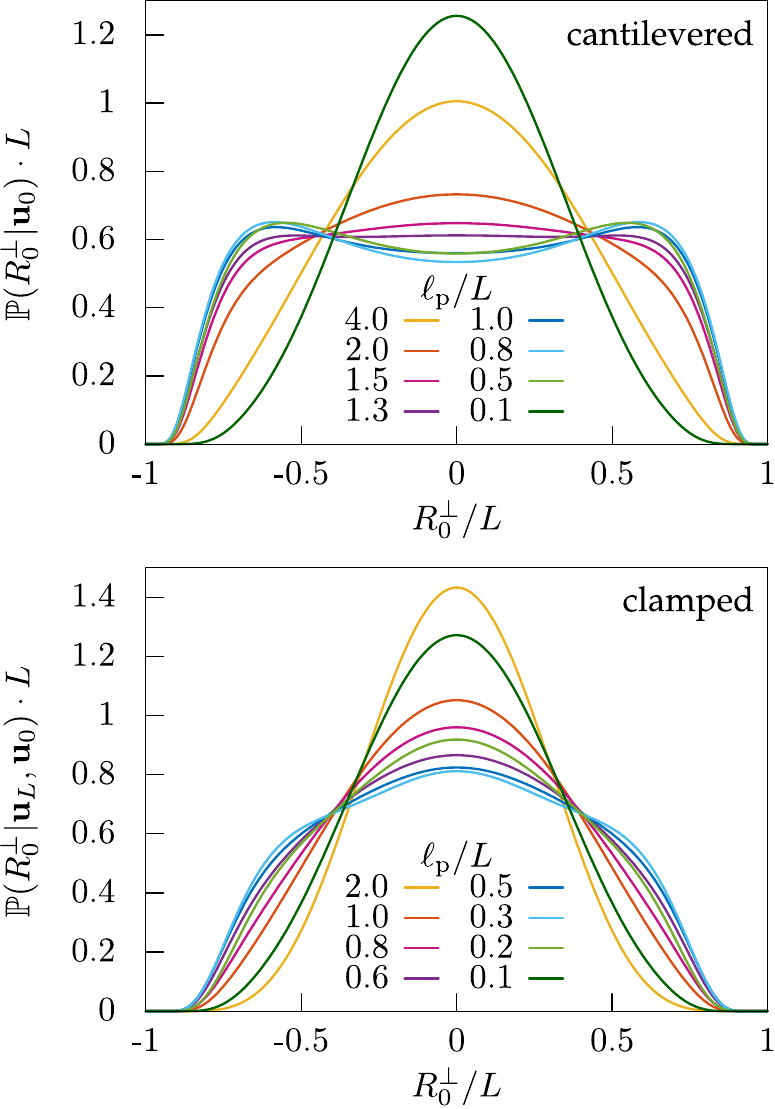}
\caption{Probability density for the end-to-end distance $R_0^\perp$ of a semiflexible polymer
with cantilevered, $\mathbb{P}(R_0^\perp|\vec{u}_0)$ (\textit{top panel}),
and clamped ends, $\mathbb{P}(R_0^\perp|\vec{u}_L,\vec{u}_0)$ (\textit{bottom panel}),
projected onto the transverse direction of the initial (and final) orientation $\vec{u}_0$($=\vec{u}_L$).
\label{fig:prob_density_hc_trans}}
\end{figure}

Transverse fluctuations of the free end of a cantilevered polymer are quantified in terms
of the probability density for the end-to-end distance projected perpendicular to the
clamped end $R_0^\perp$, see Fig.~\ref{fig:prob_density_hc_trans} (\textit{top panel}). The probability
densities for polymers with different bending rigidities display an interesting reentrant behavior,
such that for flexible, $\ell_\text{p}/L\ll1$, and for stiff polymers, $\ell_\text{p}/L\gg1$,
the probability density exhibits a single peak centered at zero projected end-to-end distance, $R_0^\perp/L=0$.
Thus, the free end of a stiff cantilevered polymer prefers to align along the same direction
as the clamped end.
\begin{figure*}[t]
\centering
\includegraphics[width = \linewidth]{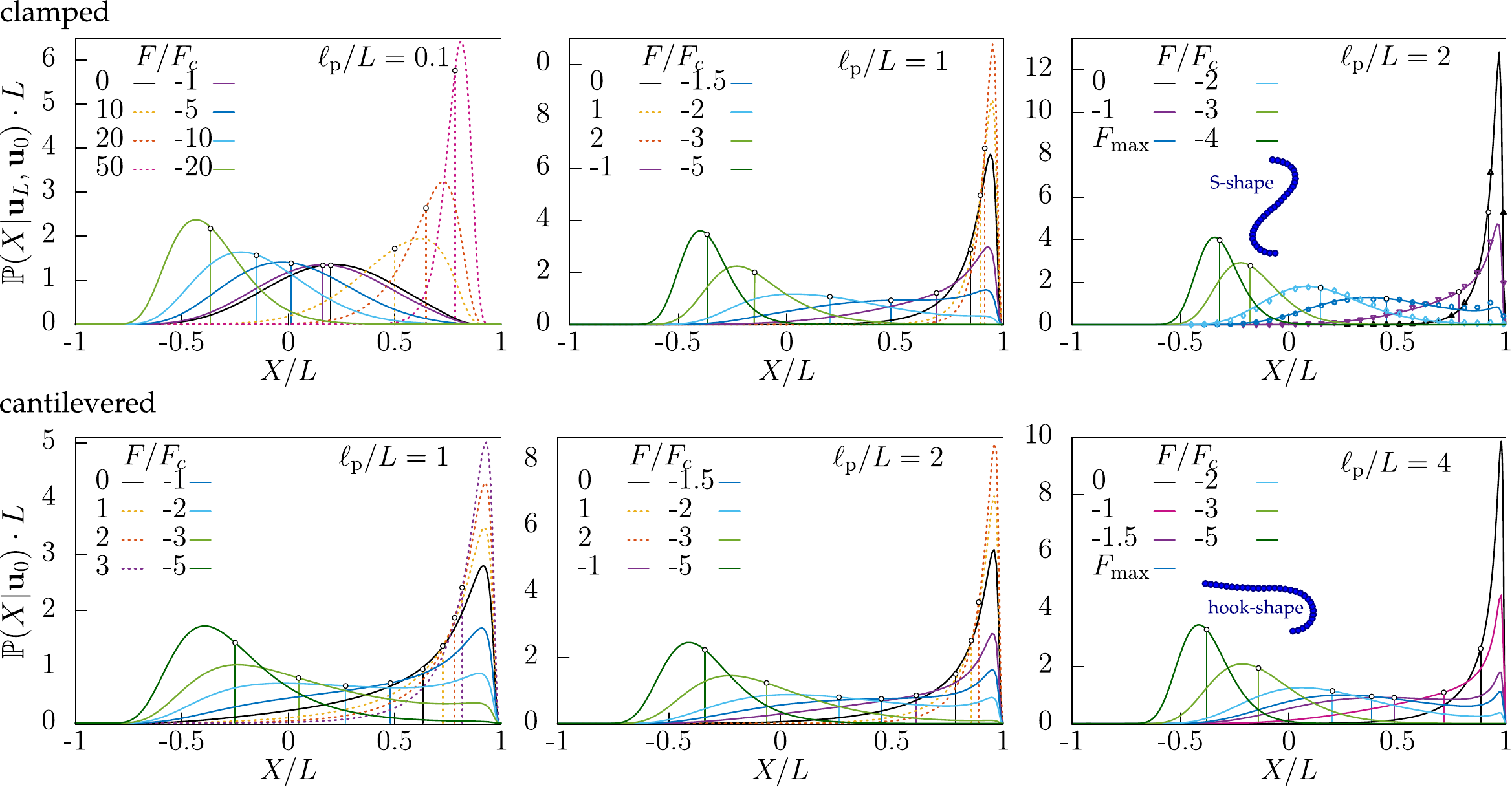}
\caption{Probability density of semiflexible polymers with clamped ends $\vec{u}_0=\vec{u}_L$,
$\mathbb{P}(X|\vec{u}_L,\vec{u}_0)$ (\textit{top row}),
and cantilevered ends, $\mathbb{P}(X|\vec{u}_0)$ (\textit{bottom row}), for the end-to-end distance $X$ projected along the
applied force, $\vec{F}=F\vec{u}_0$.
Figures show semiflexible polymers of different persistence lengths $\ell_\text{p}$ subject to external pulling (dashed lines) and compression (solid lines) forces $F$.
Here, $L$ denotes the contour length of the polymer, $F_c=\pi^2\kappa/(\gamma L)^2$ is the critical Euler buckling force with $\gamma=1$ for clamped
and $\gamma = 2$ for cantilevered polymers, and $F_\text{max}$ is the force at maximal susceptibility $\chi$.
Vertical lines with circles indicate the projected mean end-to-end distance $\langle X\rangle /L$ extracted
from the force-extension relations~\cite{Kurzthaler:2017}.
Selected pseudo-dynamic simulations are indicated by symbols.
\label{fig:prob_density_force}}
\end{figure*}
In the case of a flexible polymer, the boundary conditions can be neglected
and the probability densities for the end-to-end distance of a free polymer follow a Gaussian [Sec.~\ref{sec:force-free_free}].
As an intriguing fingerprint of the semiflexibility of the polymer, $\ell_\text{p}/L\simeq1$,
the probability density is characterized by a bimodal shape indicating the transverse fluctuations of
the free end with respect to the clamped end. This peculiar behavior arising in semiflexible polymers
has already been observed earlier for polymers in 2D
in terms of computer simulations~\cite{Lattanzi:2004} and an approximate theory~\cite{Benetatos:2005}. Similar behavior has been
observed in the distribution function of a cantilevered polymer in 3D, which has been elaborated formally exactly in Fourier-Laplace space
by an expansion in Legendre polynomials~\cite{Semeriyanov:2007}.

Qualitatively different behavior is observed for the transverse fluctuations of a polymer with
two clamped ends [Fig.~\ref{fig:prob_density_hc_trans} (\textit{bottom panel})].
Here, the probability densities display a prominent peak at zero transverse end-to-end distance, $R_0^\perp/L=0$, for all
bending rigidities $\ell_\text{p}/L$. Yet, in the regime of semiflexible polymers, $\ell_\text{p}/L\simeq1$, the probability
densities additionally exhibit two heavy tails, reflecting the transverse fluctuations between the clamped ends.

\section{Results -- semiflexible polymer subject to external loads}
In striking contrast to the buckling behavior of a classical rigid rod, which does not yield at all
but starts to buckle at the critical Euler buckling force, thermal fluctuations smear out the Euler buckling
instability for semiflexible polymers~\cite{Kurzthaler:2017}. Yet, a peak in the susceptibility of a semiflexible polymer
is still observed close to the critical Euler buckling force and, thus, we refer to this behavior as the
\textit{buckling transition} of a semiflexible polymer.
In particular, the smearing of the Euler buckling instability results in a rapidly decaying force-exension relation
with a broad variance at the buckling transition, which is reminiscent of a smeared discontinuous phase transition
in a finite system. Moreover, these lowest-order moments indicate a non-Gaussian probability density of the polymer at the bucking transition
and, therefore, a profound analysis of the full probability density is required to adequately interpret these findings.

\begin{figure*}[t]
\centering
\includegraphics[width = \linewidth]{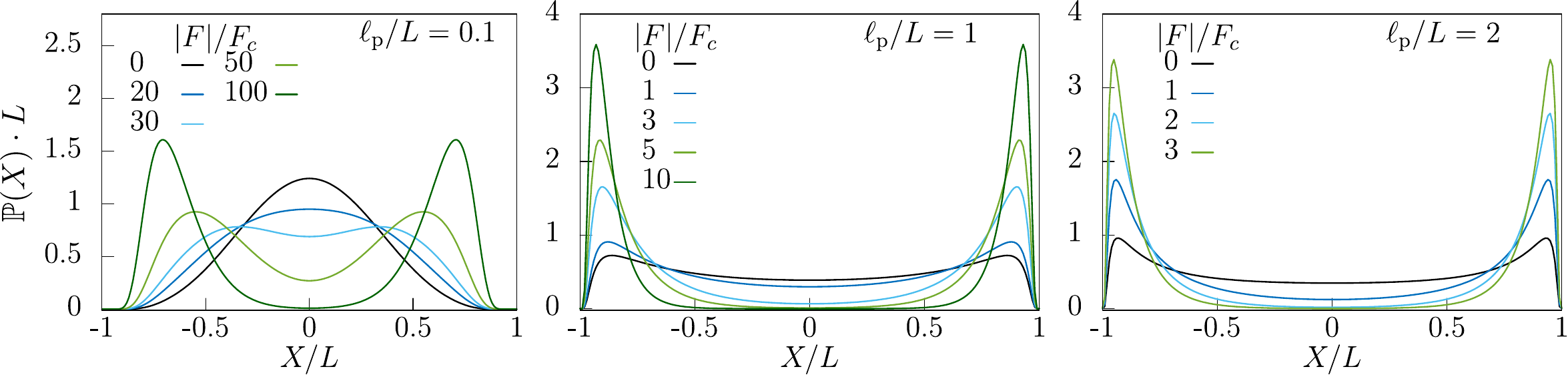}
\caption{Intersection of the circularly symmetric probability density $\mathbb{P}(X)$ for the end-to-end distance $X$
of semiflexible polymers with free ends and different persistence lengths $\ell_\text{p}$
subject to external forces $|F|$.
Here, $L$ denotes the contour length of the polymer and
the critical Euler buckling force $F_c=\pi^2\kappa/(\gamma L)^2$ with $\gamma = 2$ is used to normalize the forces.
\label{fig:prob_density_force_f}}
\end{figure*}

Here, we analyze the response of semiflexible polymers to external forces in terms of the
probability densities for their projected end-to-end distance, which permit
to shed light on the polymer configurations in the presence of a compression or
elongation force. In particular, we discuss polymers with different bending rigidities
and boundary conditions, see Fig.~\ref{fig:prob_density_force} for clamped  and cantilevered polymers and
Fig.~\ref{fig:prob_density_force_f} for free polymers.

\subsection{Clamped and cantilevered polymers}
As predicted from the force-free behavior of clamped and cantilevered polymers, the probability density for the end-to-end distance
of stiff polymers displays a left-skewed peak close to a fully stretched configuration, $X/L=1$,
whereas more flexible polymers have a broad probability density which approaches a Gaussian for
large flexibility. Under extensional forces these polymers approach almost their full
contour length with increasing forces, yet, stronger forces are required to elongate more flexible polymers,
than stiffer polymers, see Fig.~\ref{fig:prob_density_force} dashed lines. For increasing
forces the probability densities shift towards full extension $X/L=1$, where they narrow significantly.
The mean end-to-end distance projected onto the direction of the applied force $\langle X\rangle/L$
can be obtained by directly averaging the probability distribution or from the previously calculated
exact force-extentsion relation~\cite{Kurzthaler:2017}. We have checked that both approaches yield the same result
corroborating our methods.
Fig.~\ref{fig:prob_density_force} reveals that the most probable configuration is even longer than the mean end-to-end distance,
however, the heavy tail of the distribution also allows for more coil-like structured configurations
due to thermal fluctuations. Similar behaviors are observed for clamped and cantilevered polymers, respectively.

Although the stretching behavior of these clamped and cantilevered polymers remains similar with respect to their stiffness,
flexible polymers do not display a \textit{buckling transition} anymore.
For flexible polymers, the probability density assumes a unimodal shape for all applied forces;  however, the peak of the
probability density shifts towards negative end-to-end distances in the presence of compression
forces, where the probability density assumes
a right-skewed form that sharpens with increasing forces [Fig.~\ref{fig:prob_density_force} (\textit{left panel, top})].

In striking contrast, stiffer polymers display a bimodal distribution at forces in the vicinity of the
critical force $F_\text{max}$, where the susceptibility, $\chi = \left(\partial\langle X\rangle/\partial F\right)_T$,
defined as the derivative of the mean end-to-end distance $\langle X\rangle$ with respect to the applied force $F$,
exhibits a maximum; compare, for example, Fig.~2 (d) in Ref.~\cite{Kurzthaler:2017}
and Fig.~\ref{fig:prob_density_force} (\textit{right panel, top}).
The bimodal distribution reflects two favored configurations for clamped and cantilevered polymers under compression.
One configuration constitutes an almost elongated configuration, which resists the applied compression force, and
the second configuration reflects the bending of the polymer as response to the compression. In the buckled state,
clamped polymers  exhibit an S-shaped configuration with both ends aligned against the applied force
(see inset in Fig.~\ref{fig:prob_density_force} (\textit{right panel, top})).
Similarly, cantilevered polymers assume a hook-shaped configuration with the free end aligned
along the applied force (see inset in Fig.~\ref{fig:prob_density_force} (\textit{right panel, bottom})).
The corresponding mean end-to-end distance $\langle X\rangle/L$~\cite{Kurzthaler:2017} is located between the two peaks and, thus,
does not reflect the favored configuration of the polymer.
Moreover, these two modes become more prominent for polymers of increasing stiffness, which allows for a
restricted number of chain configurations only, whereas they
are smeared out for more flexible polymers due to thermal fluctuations.
For larger forces, the S-shaped (for clamped) and the hook-shaped (for cantilevered) configurations take over and
the probability density exhibits a single peak located at a negative end-to-end distance.
Interestingly, a bimodal shape is also observed in the probability density for the order parameter of a finite system
at a smeared discontinuous phase transition~\cite{Binder:1992}. In particular, the probability density
dispays two peaks at the transition temperature, whereas one mode disappears upon in/decreasing the temperature.

We have validated our analytical results by comparing the mean projected end-to-end distance $\langle X\rangle$ and the
standard deviation $\langle (X-\langle X\rangle)^2\rangle$ obtained numerically from the probability distribution to the
force extension relation and the susceptibility $\chi = \langle (X-\langle X\rangle)^2\rangle/k_\text{B}T$ elaborated exactly
in Ref.~\cite{Kurzthaler:2017}.
In addition, we have corroborated the exact probability densities for selected parameter sets with
pseudo-dynamic simulations indicated by symbols in Fig.\ref{fig:prob_density_force}
(see appendix~\ref{sec:appendix_simulations} for simulation details).

\subsection{Free polymer}
The behavior of free polymers under stretching and compression is identical, as the ends are free to align into
the direction of the applied force.
In particular, by integrating over the initial orientations
the corresponding probability densities of the end-to-end distance become circularly symmetric, and therefore, it suffices to
consider a projection along a specific direction only.
For rather stiff force-free polymers we find a probability density with two peaks located in the vicinity of $X/L=\pm 1$, which become
larger for increasing stiffness, see Fig.~\ref{fig:prob_density_force_f} for $\ell_\text{p}/L=1,2$.
Note, that the negative end-to-end distance
corresponds to polymers aligned along the opposite direction than the projection vector.
Applying an external force ((anti-) parallel to the projection vector), the
distributions become more narrow and the modes approach the fully elongated configuration.
In the case of more flexible polymers the two peaks in the force-free distribution vanish
and the probability density displays only a single peak at $X=0$ [Fig.~\ref{fig:prob_density_force_f}, $\ell_\text{p}/L=0.1$].
In the presence of an external force
two peaks emerge and approach the vicinity of $X/L=\pm 1$ for increasing forces. Here, the
forces required to fully stretch the polymer are much larger than for stiffer polymers.
Thus, as for clamped and cantilevered polymers, thermal
fluctuations allow flexible polymers to resist more strongly, than stiffer polymers, which is reflected in broader
distributions.

\section{Summary and conclusion}
We have derived exact solutions for the probability densities
for the (projected) end-to-end distance of a wormlike chain
in the presence of external forces and compared them to the force-free case.
Our analytic results have permitted to elucidate the stretching behavior and the \textit{buckling transition} of semiflexible polymers
with arbitrary bending rigidities for different experimental setups.
As most prominent feature we have found a bimodal probability density for the projected end-to-end distance
in the vicinity of the \textit{buckling transtion}, which reflects the stretched and S- (or hook-) shaped configurations of
clamped (or cantilevered) polymers induced by the compression.
In particular, we have observed a passage from a unimodal distribution at small forces reflecting the
stretched configuration of the polymer, to a bimodal distribution at the buckling transition, and again to a unimodal
distribution for the S- (or hook-) shaped configuration at large forces. Interestingly, this behavior is reminiscent of a discontinuous phase transition
in a finite system, where the probability density for the order parameter exhibits two modes
close to the transition temperature, which reduce to one upon in/decrasing temperatures~\cite{Binder:1992}.
In this sense, increasing the stiffness  and thereby suppressing the thermal fluctuations of the semiflexible polymer, plays
the same role as the infinite volume limit close to a discontinuous phase transition.
Moreover, the large variance of the force-extension relation at the
buckling transition~\cite{Kurzthaler:2017} serves to a large extend as a measure for the distance between the two favored configurations,
in contrast to the width of a unimodal distribution.
Hence, the full probability density has provided essential information for the correct interpretation of these low-order moments,
as is fundamental to understand non-Gaussian behavior.

However, these features are only present in the case of
semiflexible polymers $\ell_\text{p}/L\simeq 1$, whereas flexible polymers $\ell_\text{p}/L\ll 1$ respond
qualitatively differently to external forces.
In particular, the probability densities for flexible polymers display only one mode independent of the
applied force, which becomes broader for increasing flexibility. Thus, flexible polymers can resist
external forces more effectively due to the presence of strong thermal fluctuations.

Force spectroscopy techniques have already been applied successfully to elucidate the nonlinear elastic response of a variety of biopolymers
in terms of force-extension relations. In these experiments, one end of the polymer is
usually tethered to a surface and a colloidal bead is attached to the other end, which reflects the setup
of a cantilevered polymer. Using, for example,
magnetic\cite{Gosse:2002} or optical tweezers~\cite{Mehta:1999} the colloid can be trapped, moved and steered,
thereby, exerting a force on the attached polymer.
Although experimental observations are mostly restricted to pulling forces, also compression forces could be applied in a similar manner.
Moreover, tracking the bead should in principle permit to extract reliable information on the probability density
for the end-to-end distance of the polymer. Thus, we anticipate, that a direct comparison of experiments to our
theoretical predictions allows deepening the understanding of the elastic behavior of semiflexible polymers
in the presence of external forces.

Further theoretical investigations are needed to unravel the elastic properties of different types of polymers, including,
for instance polymers in 3D or curved polymers. Our solution strategy can be transferred to semiflexible
polymers in 3D, where the Mathieu functions are replaced by the generalized spheroidal wave functions, as
elaborated for the mathematical analog of an active Brownian particle in bulk~\cite{Kurzthaler:2016}. Similarly,
analytical progress can be achieved by accounting for a spotaneous curvature of the polymer configuration such
that the classical reference system displays a circular arc~\cite{Potemkin:2004}. In this case the mathematical analog is the Brownian
circle swimmer and the eigenfunctions become generalizations of Mathieu functions~\cite{Kurzthaler:2017:circle}.

In contrast to an external force acting on the end of the polymer, external fields, including, for examlpe,
hydrodynamic flows~\cite{Perkins:1995, Larson:1997, Smith:1998} or electric fields~\cite{Ferree:2003,Li:2012}, can be used for
the manipulation of single polymers. Theoretical progress~\cite{Lamura:2001,Benetatos:2004,Manca:2012} has
been made by considering a tethered polymer in an external field, where a force acts on each bead of the discretized polymer chain.
Similar to the behavior of a polymer under extension, the tethered polymer aligns along the direction of the external field and its mean
configuration is characterized by the orientation of the final segment and its presistence length~\cite{Lamura:2001}.
In the presence of an additional external force applied opposite to the direction of the external field,
the polymer displays a hook-shaped configuration~\cite{Manca:2012}, analogous to a cantilevered polymer under compression.
Interestingly, charged polymers display even more complex behavior in an electric field, as
the charge distribution determines the direction of the polymer deformations~\cite{Benetatos:2004}. In particular,
an electric field can also compress the polymer and the elastic behavior should be qualitatively similar to the \textit{buckling transition} of
a wormlike chain.

In addition to the elastic properties of single polymers, the behavior of single polymers inside cells
and the buckling behavior of entire networks composed of semiflexible polymers remains to be fully understood.
We anticipate, that the single-polymer behavior in free space, as elaborated here,
constitutes the reference for the more realistic case
of polymers immersed in a densely crowded environment~\cite{Keshavarz:2016,Leitmann:2016, Schobl:2014}
and moreover, serves as input for the analysis of polymer networks~\cite{Amuasi:2015,MacKintosh:2014,Carillo:2013,Chaudhuri:2007,Claessens:2006,
Huisman:2008,Kroy:1996,MacKintosh:1995,Plagge:2016,Razbin:2015,Storm:2005}.
In a simplified description, these polymeric networks can be regarded as entangled solutions
of semiflexible polymers, which can cross or loop each other, exhibit branching points,
or are tightly connected by cross-links.
Consequently, a polymer naturally experiences compression or stretching forces
due to the constraints set by the surrounding environment. Thus, our predictions for the conformational properties
of single polymers induced by external forces may constitute a convenient
starting point for studying the equilibrium properties of polymer networks.

Beyond the force-free behavior of networks, the elastic response of these polymeric structures to external loads is
of fundamental interest and is substantially governed by the buckling behavior and the corresponding shapes of the single components.
In particular, our theoretical predictions for the probability densities of semiflexible
polymers of arbitrary stiffness may be useful to explore the mechanical stability of such networks in the regime of
strong compression, where the polymers display strongly pronounced S- or hook-shaped configurations and
the weakly-bending approximation breaks down.

\section*{Acknowledgments}
This work has been supported by the Austrian Science Fund (FWF): P~28687-N27.

\section*{Appendix}
\appendix

\section{Numerical evaluation of the Mathieu functions\label{sec:appendix_numerics}}
The even and odd Mathieu functions,  $\text{ce}_{2n}(q)$ [Eq.~\eqref{eq:mathieu_even}] and $\text{se}_{2n+2}(q)$ [Eq.~\eqref{eq:mathieu_odd}],
and the associated eigenvalues, $a_{2n}(q)$ and $b_{2n+2}(q)$, are evaluated numerically
by solving the recurrence relations for the Fourier coefficients~\cite{NIST:online, NIST:print, Ziener:2012}.
In particular, we obtain the recurrence relations for the Fourier coefficients of the even Mathieu functions $A_{2m}^{2n}(q)$ by
inserting Eq.~\eqref{eq:mathieu_even} into the Mathieu equation [Eq.~\eqref{eq:mathieu}],
\begin{align}
\begin{split}
  a_{2n}A_0^{2n}-qA_2^{2n}  &=0,\\
  (a_{2n}-4)A_2^{2n}-q(2A_0^{2n}+A_4^{2n})  &=0,\\
(a_{2n}-4m^2)A_{2m}^{2n}-q(A_{2m-2}^{2n}+A_{2m+2}^{2n})&=0, \quad m\geq 2,\label{eq:rec_A}
\end{split}
\end{align}
and, similarly, for the Fourier coefficients of the odd Mathieu functions
$B_{2m+2}^{2n+2}(q)$
\begin{align}
\begin{split}
  (b_{2n+2}-4)B_2^{2n+2}-qB_4^{2n+2}  &=0,\\
(b_{2n+2}-4m^2)B_{2m}^{2n+2}-q(B_{2m-2}^{2n+2}+B_{2m+2}^{2n+2})&=0, \quad m\geq 2.\label{eq:rec_B}
\end{split}
\end{align}
The Fourier coefficients are computed numerically by transforming the recurrence relations into a
matrix eigenvalue problem, $\textsf{M}^A\vec{A}^{2n}=a_{2n}\vec{A}^{2n}$ and $\textsf{M}^B\vec{B}^{2n+2}=b_{2n+2}\vec{B}^{2n+2}$,
where the eigenvectors contain the Fourier coefficients, $\vec{A}^{2n} = [\sqrt{2}A_0^{2n}, A_2^{2n}, A_4^{2n},\dots]$
and $\vec{B}^{2n} = [B_2^{2n}, B_4^{2n}, B_6^{2n},\dots]$. The matrix $\textsf{M}^A$ is a band matrix
with diagonal elements $M^A_{mm} = 4m^2$ and off-diagonal elements $M^A_{0,1}=M^A_{1,0}=\sqrt{2}q$, $M^A_{1,2}=q$,
 and for $m\geq 2$: $M^A_{m,m+1}=M^A_{m,m-1}=q$. Similarly, $\textsf{M}^B$ containes diagonal elements
$M^{B}_{0,0}=4$ and for $m\geq1$: $M^B_{mm} = 4m^2$, and off-diagonal elements $M^B_{0,1}=q$,
 and for $m\geq 1$: $M^B_{m,m+1}=M^B_{m,m-1}=q$. The orthonormalization of the Mathieu functions translates for
the Fourier coefficients to $2 A_0^{2n} A_0^{2m}+ \sum_{j\geq1} A_{2j}^{2n}A_{2j}^{2m}=\delta_{nm}$ and
$\sum_{j} B_{2j+2}^{2n+2} B_{2j+2}^{2m+2}=\delta_{nm}$~\cite{NIST:online, NIST:print, Ziener:2012}.

In practice, the square matrix is truncated at an appropriate dimension (i.e. $\text{dim} \ \textsf{M}^{A,B}\approx 100$) to achieve
the desired accuracy for the Mathieu functions. These eigenfunctions are labeled with respect to the real part of the associated eigenvalue,
$\text{Re}[a_{0}]\leq\text{Re}[a_{2}]\leq\text{Re}[a_{4}]\leq\dots$
and $\text{Re}[b_{2}]\leq\text{Re}[b_{4}]\leq\text{Re}[b_{6}]\leq\dots$.
Thus, the higher modes of the partition sums [Eqs.\eqref{eq:characteristic_polymer_mathieu},~\eqref{eq:characteristic_polymer_hc},~\eqref{eq:characteristic_polymer_free}]
become exponentially suppressed due to the increasing eigenvalues, which induces a natural cut-off of the infinite series.%~\cite{Kurzthaler:2017}.

\section{Pseudo-dynamic simulations\label{sec:appendix_simulations}}
We validate our analytical results by pseudo-dynamic simulations following the scheme
elaborated in Ref.~\cite{Kurzthaler:2017}. Here, the contour $L$ of the polymer
is discretized in terms of equidistantly separated beads $\{\vec{R}_i\}_{i=0}^{N}$ and
corresponding tangent vectors $\{\vec{u}_i\}_{i=0}^{N-1}$, where
$\vec{u}_i=(\vec{R}_{i+1}-\vec{R}_{i})N/L$ is of unit length, $|\vec{u}_i|=1$.
The time evolution of the orientation of the $i$-th bead is encoded in
the Langevin equation in the It$\bar{\text{o}}$ sense
\begin{align}
\begin{split}
   \diff & \vec{u}_i(t)   = -\hat{D}_\text{rot}\vec{u}_i(t)\diff t + \hat{D}_\text{rot}\vec{u}_i^\perp(t)\Bigl(\frac{N}{L}\frac{\ell_\text{p}}{2}\vec{u}_i^\perp(t)\cdot[\vec{u}_{i-1}(t)\\
   & +\vec{u}_{i+1}(t)]-\frac{L}{N}|f|\vec{u}^\perp_i(t)\cdot\vec{e}\Bigr)\diff t + \sqrt{2\hat{D}_\text{rot}} \vec{u}_i^\perp(t) \diff \omega_i(t),\label{eq:sim}
\end{split}
\end{align}
for $i=1,\dotsc,N-2$. The unit orientation rotated clockwise by an angle of $\pi/2$ is denoted by $\vec{u}_i^\perp(t)$ with the properties
$\vec{u}_i^\perp(t) \cdot \vec{u}_i(t)= 0$ and  $\det[\left( \vec{u}_i(t),\vec{u}_i^\perp(t)\right)]=1$.
Moreover, $\omega_i(t)$ is a Gaussian white noise process with
zero mean and delta-correltated variance $\langle\omega_i(t)\omega_j(t')\rangle=\delta_{ij}\delta(t-t')$ for
$i,j=0,\dotsc,N-1$. The time scale for the relaxation of the orientations of the segments is set by the
scaled rotational diffusion coefficient $\hat{D}_\text{rot}$.

At the ends, $i=0,N-1$, the Langevin equations are modified according to the
boundary conditions. In particular, for a clamped polymer under compression the ends are aligned into the opposite direction of the
applied force, $\vec{u}_0(t)=\vec{u}_L(t)= \vec{e}$, and similarly the clamped end of a cantilevered polymer is fixed, $\vec{u}_0(t)=\vec{e}$,
whereas its final orientation $\vec{u}_L(t)$ can freely rotate.

To corroborate our theoretical predictions for the probability densities, configurations of the polymer are extracted from the simulations after
the polymer has reached equilibrium. For the selected simulations we have obtained reliable statistics by simulating $10$ realizations of polymers
with $N=300$ segments using time steps of $10^{-5}/\hat{D}_\text{rot}$ over a time horizon
of $10^{4}/\hat{D}_\text{rot}$.


% \balance


%%%REFERENCES%%%

\bibliography{polymer_bib}
\bibliographystyle{rsc}

\end{document}